# Structural and Electronic Properties of $Ta_2O_5$ with One Formula Unit

**Yangwu Tong[1,2#] , Huimin Tang[3#], and Yong Yang[1,2,3*]**

1. *Key Lab of Photovoltaic and Energy Conservation Materials, Institute of Solid State Physics, HFIPS, Chinese Academy of Sciences, Hefei 230031, China.*
2. *Science Island Branch of Graduate School, University of Science and Technology of China, Hefei 230026, China.*
3. *College of Physics and Technology, Guangxi Normal University, Guilin 541004, China.*

**ABSTRACT:**

Based on particle swarm optimization (PSO) algorithm and density functional theory (DFT) calculations, we identify a stable triclinic crystal structure of $Ta_2O_5$ (named as $\gamma_1$-$Ta_2O_5$) at atmospheric pressure whose unit cell contains one formula unit (Z=1). Comparison with the Z=1 $Ta_2O_5$ structures from the Materials Project [*APL Mater.* 1, 011002 (2013)] reveals that $\gamma_1$-$Ta_2O_5$ is energetically the most stable among the Z=1 $Ta_2O_5$ phases, and is the second most stable among *all* the $Ta_2O_5$ phases. Characterization of $\gamma_1$-$Ta_2O_5$ is carried out by analyzing the X-ray powder diffraction patterns, the elastic, vibrational, thermal and electronic properties. The electronic structures of $\gamma_1$-$Ta_2O_5$ are calculated using standard DFT as well as many-body perturbation theory within the *GW* approximation. The results indicate that $\gamma_1$-$Ta_2O_5$ is a wide band gap semiconductor with an indirect gap of ~ 3.361 eV.

**Keywords:** $Ta_2O_5$, Structure Search, Elementary Building Block, DFT Calculations

[#]These authors contribute equally to this work.

*****Corresponding author: yyanglab@issp.ac.cn

## 1. Introduction

$Ta_2O_5$ is a wide band gap transition-metal oxide semiconductor with high dielectric constant and adjustable band gap which lead to a wide range of applications such as optical materials [1-3], dielectric materials [4,5], photocatalysis and electrocatalysis [6-8], and non-volatile resistive random access memories [9-13]. The thin films of $Ta_2O_5$ have high chemical stability, low reflectivity and high refractive index. Therefore, $Ta_2O_5$ has been used as anti-reflection layers on the surface of the silicon solar cell to reduce the reflection of sunlight and improve the efficiency of the silicon solar cell [14]. Due to the large difference of refractive index between $Ta_2O_5$ and $SiO_2$, the film formed by alternating layers of the two materials shows high reflectivity. $TiO_2$-doped $Ta_2O_5/SiO_2$ multilayer structures are employed as the mirror optical coatings of the Laser Interferometer Gravitational-wave Observatory (LIGO) [15, 16]. Special attention has been paid on $TiO_2$ (or $ZrO_2$) doped amorphous $Ta_2O_5$ which reduces the thermal noise and consequently the mechanical loss of the highly reflective mirror coating layers used in LIGO [17-21].

Recently, $Ta_2O_5$ has also been taken as a candidate of surface ceramic coating materials for artificial bones and teeth due to its excellent biocompatibility, high wear and corrosion resistance [22, 23]. The high-temperature phase of $Ta_2O_5$ undergoes spontaneous phase transitions during cooling [24, 25], which makes it difficult to grow high-quality single crystals of $Ta_2O_5$. Therefore, experimental samples of $Ta_2O_5$ prepared at room temperature usually contain some defects. Dilute oxygen vacancies in the crystal will break the charge balance with adjacent Ta atoms, which in turn induces the mobility of O atoms near the vacancy sites, and cause significant modifications to electronic properties [26, 27]. The presence of oxygen vacancies can further drive the displacement of atomic positions and significantly distort the O sublattice, which leads to short and intermediate range of structural disorder [26-28]. The results of the molecular dynamics simulations by Lee et al. [28] indicate that two-fold coordinated O atoms move widely in the vertical direction of the line connecting the two coordinating Ta atoms, which may cause them to enter the three-coordinated sites temporarily. The structural change due to the motions of two-fold coordinated O atoms may also the reason of structural disorder. Although the structures of $Ta_2O_5$ have been studied for decades, the issue regarding its ground-state crystal structures remains open. Indeed, it has been shown that long-range lattice relaxation can be induced by a charge-neutral oxygen vacancy, and the potential energy surface related to the variation of Ta-O bond lengths imply that the crystal structures at the vicinity of ground-state may be highly degenerate [27, 29]. In a recent work by one of the authors, a triclinic phase ($\gamma$-$Ta_2O_5$) whose unit cell contains two formula units (Z=2) is predicted [30] and is found to be energetically more stable than any phases/models of $Ta_2O_5$ reported previously. Soon afterwards, it is shown that the supercells of $\gamma$-$Ta_2O_5$ intrinsically contain a high symmetry phase (Z=4) with the space group of $I4_1/amd$ [31].

In this work, we study the structural and electronic properties of a number of Z= 1 $Ta_2O_5$ phases which may be taken as the elementary building block of all other the phases whose unit cells contain more formula units (Z≥2). In particular, we will focus on a triclinic phase obtained through extensive structure search. This phase is shown to be energetically the most stable among the Z=1 $Ta_2O_5$ structures. An in-depth study on the vibrational, thermodynamic, elastic and electronic properties has been made using first-principles calculations. Our work reveals that this phase corresponds to a wide band gap semiconductor with an indirect band gap of ~ 3.361 eV.

## 2. Methodology

First-principles calculations based on DFT have been carried out using Vienna *ab initio* Simulation Package (VASP) [32, 33], which is interfaced with the Crystal Structure Analysis by Particle Swarm Optimization (CALYPSO) code [34] for doing structure search. CALYPSO has demonstrated its reliability in predicting thermodynamically stable new phases of materials [35-37]. In this work, the structural evolution is performed in $Ta_2O_5$ systems whose unit cells contain one formula unit (Z=1). For each generation, 50 candidate structures are produced. All structures of the first generation are randomly generated. Each subsequent generation produces 30 structures by PSO, and the other 20 structures are randomly generated with symmetry constraints. The projector augmented wave (PAW) potentials [38, 39] are employed to describe the ion (core)-electron interactions, with an energy cutoff of 600 eV for the plane wave basis set of electron wave functions. The Perdew-Burke-Ernzerhof (PBE) [40] type generalized gradient approximation (GGA) is used to describe the exchange-correlation interactions of electrons. The Brillouin zone (BZ) is sampled with a k-point spacing of ~ $0.06\times 2\pi$ Å$^{-1}$. The total energy of each structure converges to a level of less than 0.01 eV per unit cell by preliminary structural relaxation. After structure search, we optimized the lowest-energy structure with a dense $8 \times 8 \times 6$ Monkhorst-Pack k-mesh [41] to have the total energy converge to within a threshold of $1\times 10^{-6}$ eV per unit cell.

To analyze the properties of the obtained structures, the phonon dispersion and vibrational density of states (VDOS) are calculated based on density functional perturbation theory (DFPT) [42]. We have further optimized the unit cell so that the total energy and force converge to within a threshold of $1\times 10^{-10}$ eV and $3\times 10^{-4}$ eV/Å, respectively. The Hessian matrix is obtained by DFPT calculation using a $2 \times 2 \times 2$ supercell. The force constants, phonon dispersion and VDOS are then obtained with the aid of the PHONOPY package [43].

To accurately evaluate the band gap, we employ the *GW* method [44, 45], which explicitly includes the many-body effects of interacting electrons to calculate the energy levels of low-lying excited states and therefore the band gap. Specially, we take the one-shot *GW* method implemented in VASP [46] to get the energy spectrum. The quasiparticle energies are obtained by solving the following equation [45]:

$$(T + V_{ext} + V_H)\psi_{nk}(\vec{r}) + \int d\vec{r}\,'\Sigma(\vec{r},\vec{r}\,'; E_{nk})\psi_{nk}(\vec{r}\,') = E_{nk}\psi_{nk}(\vec{r}), \qquad (1)$$

where $T$ is the kinetic energy operator of electrons, $V_{ext}$ is the external potential due to ions, $V_H$ is the electrostatic Hartree potential, $\Sigma$ is the electron self-energy operator, and $E_{nk}$ and $\psi_{nk}(\vec{r})$ are the quasiparticle energies and wave functions, respectively. Within the *GW* approximation [44], the term $\Sigma$ may be calculated as follows:

$$\Sigma(\vec{r},\vec{r}';E) = \frac{i}{2\pi}\int d\omega e^{i\delta\omega} G(\vec{r},\vec{r}';E+\omega)\, W(\vec{r},\vec{r}';\omega), \qquad (2)$$

where $G$ is the Green's function, $W$ is the dynamically screened Coulomb interaction, and δ is a positive infinitesimal. In our $GW$ calculations, a 6×6×4 k-mesh is used to sample the BZ, and the number of energy bands involved in the calculation is 150. This set of parameters ensures the $GW$ band gaps converge to within an error bar of ~ 0.1 eV.

## 3. Results and Discussion

### 3.1. Structural Characterization and Elastic properties of $\gamma_1$-$Ta_2O_5$

Shown in Fig. 1(a), is the ground-state energy of $Ta_2O_5$ as a function of the generation number during structure search, which drops quickly after the fifth generation. The energy is well converged after 25 generations of structural evolution. The obtained $Ta_2O_5$ structure belongs to triclinic crystal system with a space group of P1. For simplicity, we refer to this structure as $\gamma_1$ phase in the following. Its unit cell parameters are as follows: $a$ = 3.87 Å, $b$ = 3.91 Å, $c$ = 6.82 Å, α = 90.32°, β = 73.61 °, γ = 90.00 °. The atomic coordinates and Wyckoff positions of the $\gamma_1$ phase are presented in Table 1.

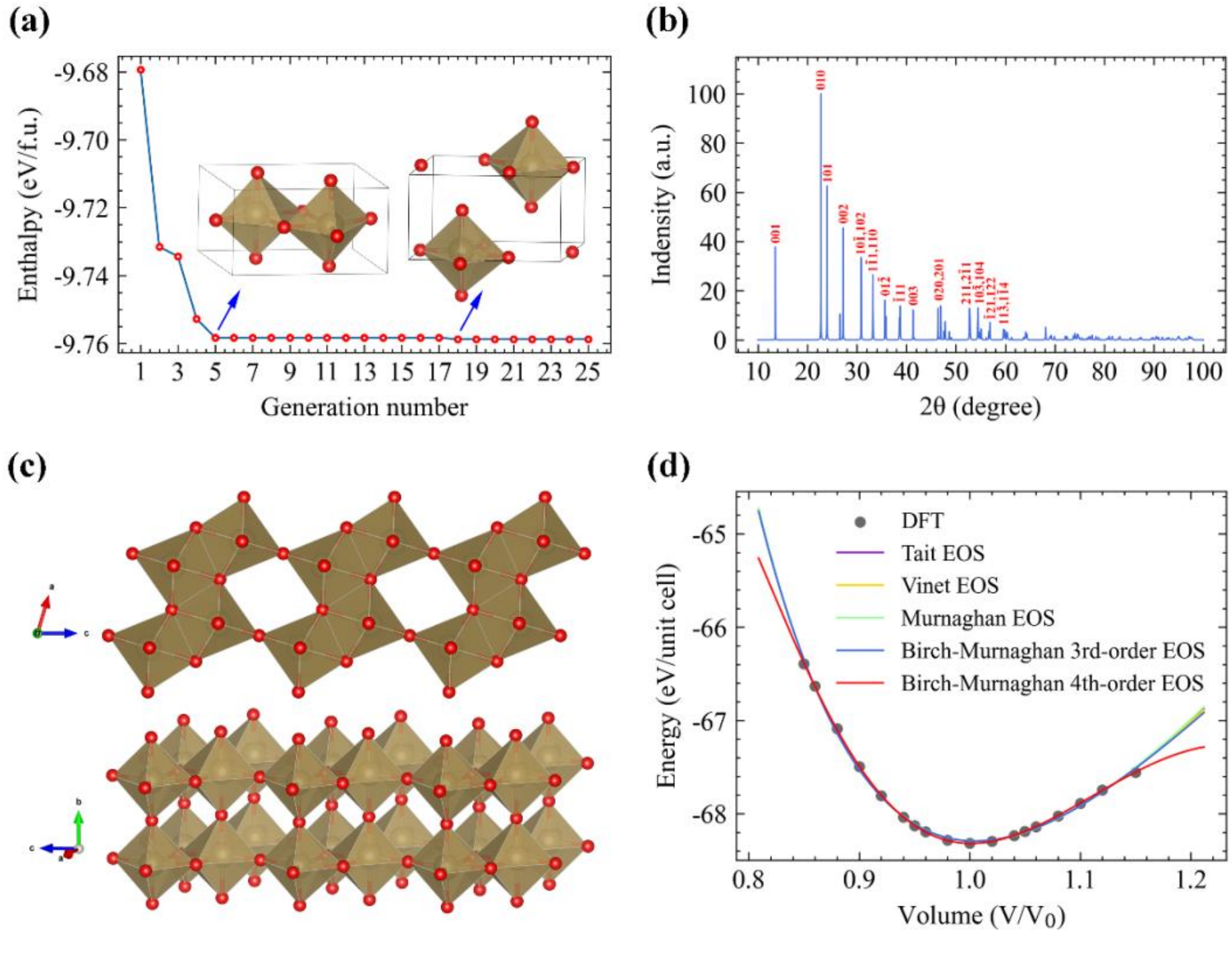

**Fig. 1.** Structure search and characterization of $\gamma_1$-$Ta_2O_5$. (a) Enthalpy evolution of the Z=1 $Ta_2O_5$ structure at atmospheric pressure as a function of the generation number predicted by PSO. (b) Simulated X-ray diffraction (Cu $K\alpha1$ radiation, wavelength ~ 1.541 Å) pattern. (c) Schematic structural model of $\gamma_1$-$Ta_2O_5$. The large (brown) and small (red) balls represent Ta and O atoms, respectively. (d) Total energy as a function of volume (normalized to the equilibrium volume at $P$ = 0). The discrete points are the data calculated by DFT at different volumes. The data are fitted according to the Tait, Vinet, Murnaghan, Birch-Murnaghan 3rd-order and 4th-order equations of state (EOS).

The optimized lattice and energetic parameters of $\gamma_1$-$Ta_2O_5$ are listed in Table 1 along with some other phases reported previously, including $\gamma$ [30], B [47, 48], $\lambda$ [49], $L_{SR}$ [50], $\delta$ [51], and $\beta_{AL}$ [52]. Additionally, three Z=1 $Ta_2O_5$ crystal phases from the Materials Project (MP) [53] (labeled as MP-1, MP-2 and MP-3) are also presented in Table 2. Compared with the original (unrelaxed) phases from the MP [53] and the other previously known phases, it can be seen that $\gamma_1$ phase is the second most stable structure (after $\gamma$ phase) and is the most stable phase among the Z=1 $Ta_2O_5$ structures. We have further optimized the three MP structures (with the fully optimized structures labeled as MP-1_new, MP-2_new and MP-3_new, respectively) using DFT calculations. Compared to the original MP-1 phase, the lattice parameters of MP-1_new change greatly, indicating that the original MP-1 phase from the MP database is unstable. Judging from their unit cell parameters, their ground-state energies (with reference to $\gamma$ phase [30]), and their band gap data (Table 2), the MP-1_new and MP-2_new phase and the $\gamma_1$ phase are actually *the same structure* within the computational error bar. Moreover, it can also be seen from Table 2 that the MP-3 (and MP-3_new) phase is actually isomorphic to the $\beta_{AL}$ phase (Z=2) found by experiment [52]: The $c$-axis of the $\beta_{AL}$ is two times that of the $b$-axis of MP-3 phase, and the $a$- and $b$-axis of $\beta_{AL}$ phase are actually the $c$- and $a$-axis of MP-3 phase. Finally, the structure dependence of the band gap is clearly seen. The variation of band gap with $Ta_2O_5$ polymorphs is in line with previous calculations [54]. From the large difference (from ~ 0.2 eV to 2.2 eV) in the band gaps ($E_g$) of Z=1 phases listed in Table 2, one sees the possibility of tuning the band gap and consequently the optical properties of polycrystalline $Ta_2O_5$ via the mixing of different phases.

**Table 1.** The fractional coordinates and Wyckoff positions of the atoms in $\gamma_1$-$Ta_2O_5$.

| Atom | *x* | *y* | *z* | Wyckoff position |
|---|---|---|---|---|
| Ta1 | 0.35326 | 0.02934 | 0.29230 | 2i |
| Ta2 | 0.64674 | 0.97066 | 0.70770 | 2i |
| O1 | 0.81414 | 0.00145 | 0.37071 | 2i |
| O2 | 0.35059 | 0.50745 | 0.29664 | 2i |
| O3 | 0.18586 | 0.99855 | 0.62929 | 2i |
| O4 | 0.64941 | 0.49255 | 0.70336 | 2i |
| O5 | 0.50000 | 0.00000 | 0.00000 | 1d |

**Table 2.** Calculated lattice and energetic parameters of $\gamma_1$-$Ta_2O_5$ and the other Z=1 $Ta_2O_5$ structures from the Materials Project (MP) database, where MP-1, MP-2, and MP-3 are respectively the No. mp-1218141, No. mp-1238961, and No. mp-554867 of MP, which are fully optimized with DFT calculations using the same computational accuracy as that for $\gamma_1$-$Ta_2O_5$. The parameters $a$, $b$, $c$ and $\alpha$, $\beta$, $\gamma$ are the axial lengths and axis-axis intersection angles of the unit cell. $\rho$ is the mass density and $\Delta E$ is the ground-state energy difference (per formula unit) with respect to the γ phase. $E_g$ is the GGA gap obtained in this work, and the band gap data of MP are presented in parentheses. The results are compared with that of γ, B, λ, $L_{SR}$, δ, and $\beta_{AL}$ phase.

| $Ta_2O_5$ phases | *a* (Å) | *b* (Å) | *c* (Å) | *α* (°) | *β* (°) | *γ* (°) | *ρ* (g/cm$^3$) | Δ*E* (eV/f.u.) | $E_g$ (eV) |
|---|---|---|---|---|---|---|---|---|---|
| $\gamma_1$ | 3.87 | 3.91 | 6.82 | 90.32 | 73.61 | 90.00 | 7.41 | 0.088 | 2.27 |
| MP-1 | 3.40 | 3.86 | 8.21 | 90.75 | 94.89 | 90.12 | 6.84 | 1.151 | 2.54(2.59) |
| MP-1_new | 3.87 | 3.89 | 6.79 | 90.09 | 104.96 | 90.03 | 7.43 | 0.098 | 2.11 |
| MP-2 | 3.84 | 3.86 | 6.78 | 90.00 | 105.57 | 90.00 | 7.58 | 0.129 | 2.18(2.34) |
| MP-2_new | 3.87 | 3.89 | 6.82 | 90.00 | 105.64 | 90.00 | 7.43 | 0.095 | 2.11 |
| MP-3 | 3.70 | 3.89 | 6.53 | 90.00 | 90.00 | 90.00 | 7.81 | 2.354 | 0.18(0.26) |
| MP-3_new | 3.69 | 3.89 | 6.52 | 90.00 | 90.00 | 90.00 | 7.85 | 2.350 | 0.21 |
| γ | 3.89 | 3.89 | 13.38 | 81.77 | 98.25 | 89.67 | 7.42 | 0 | 2.30 |
| B | 12.93 | 4.92 | 5.59 | 90.00 | 103.23 | 90.00 | 8.48 | 0.117 | 3.06 |
| λ | 6.25 | 7.40 | 3.82 | 90.00 | 90.00 | 90.00 | 8.29 | 0.243 | 2.09 |
| $L_{SR}$ | 6.33 | 40.92 | 3.85 | 90.00 | 90.00 | 89.16 | 8.10 | 0.309 | 1.92 |
| δ | 7.33 | 7.33 | 3.89 | 90.00 | 90.00 | 120.00 | 8.12 | 1.996 | 1.07 |
| $\beta_{AL}$ | 6.52 | 3.69 | 7.78 | 90.00 | 90.00 | 90.00 | 7.85 | 2.326 | 0.20 |

The simulated X-ray diffraction pattern and structural model of $\gamma_1$-$Ta_2O_5$ are shown in Figs. 1(b) and 1(c), respectively. According to the X-ray diffraction patterns, the strongest diffraction signal appears at $2\theta = 22.823°$, which corresponds to (010) diffraction plane, with a *d* spacing of 3.89 Å. The smallest diffraction angle is located at $2\theta = 13.50°$, whose diffraction plane is (001), with a *d* spacing of 3.28 Å. In the extended crystal model consisting of several unit cells (Fig. 1(c)), it is seen that each Ta atom is coordinated by six O atoms, forming an octahedral coordination structure ($TaO_6$) with the Ta atom sitting at the center and the O atoms at the vertexes. Viewing along *b*-axis, one finds a stack of coordinating octahedral layers. Two neighboring octahedra in adjacent layers are connected in the form of a shared O atom at the vertex site. For each $TaO_6$ octahedron, the number of O atoms bonded with three Ta atoms and that bonded with two Ta atoms are both 3. In a unit cell, the number of O atoms bonded with 3 Ta atoms and 2 Ta atoms are 2 and 3, respectively. The average Ta-O bond length is 2.008 Å and the polyhedral volume is 10.385 Å$^3$.

To show the structural characteristics of $\gamma_1$-$Ta_2O_5$, we have made a systematic analysis on geometries of the $Ta_2O_5$ phases listed in Table 2, with focus on the space packing of $TaO_n$ polygons, the elementary building block of $Ta_2O_5$. As seen from Fig.

2, most of the $Ta_2O_5$ phases can be constructed from the quasi-two-dimensional (2D) patterns of $TaO_n$ polygons with the basal plane (determined by $TaO_{n-2}$) aligned in the same normal direction. The only exception is B-$Ta_2O_5$, a phase obtained via high-pressure synthesis [47, 48], in which the orientations of $TaO_6$ polygons are disordered which may be due to pressure induced collapse of the original quasi-2D patterns. Table 3 compares the numbers of $TaO_n$ polygons (defined by Ta-O bond lengths), the neighboring $TaO_n$ polygons surrounding each polygon (defined by Ta-Ta distances), and the volumes of the $TaO_n$ polygons. It is found that the phases ($\gamma_1$, $\gamma$, B, $\lambda$) of high stability share the common feature that they all consist of identical $TaO_6$ octahedrons. Furthermore, as indicated in Fig. 2, the local arrangement of $TaO_6$ in $\gamma_1$ resembles one of the unique features of $\gamma$ phase: The existence of two side-by-side $TaO_6$ octahedrons in the quasi-2D patterns.

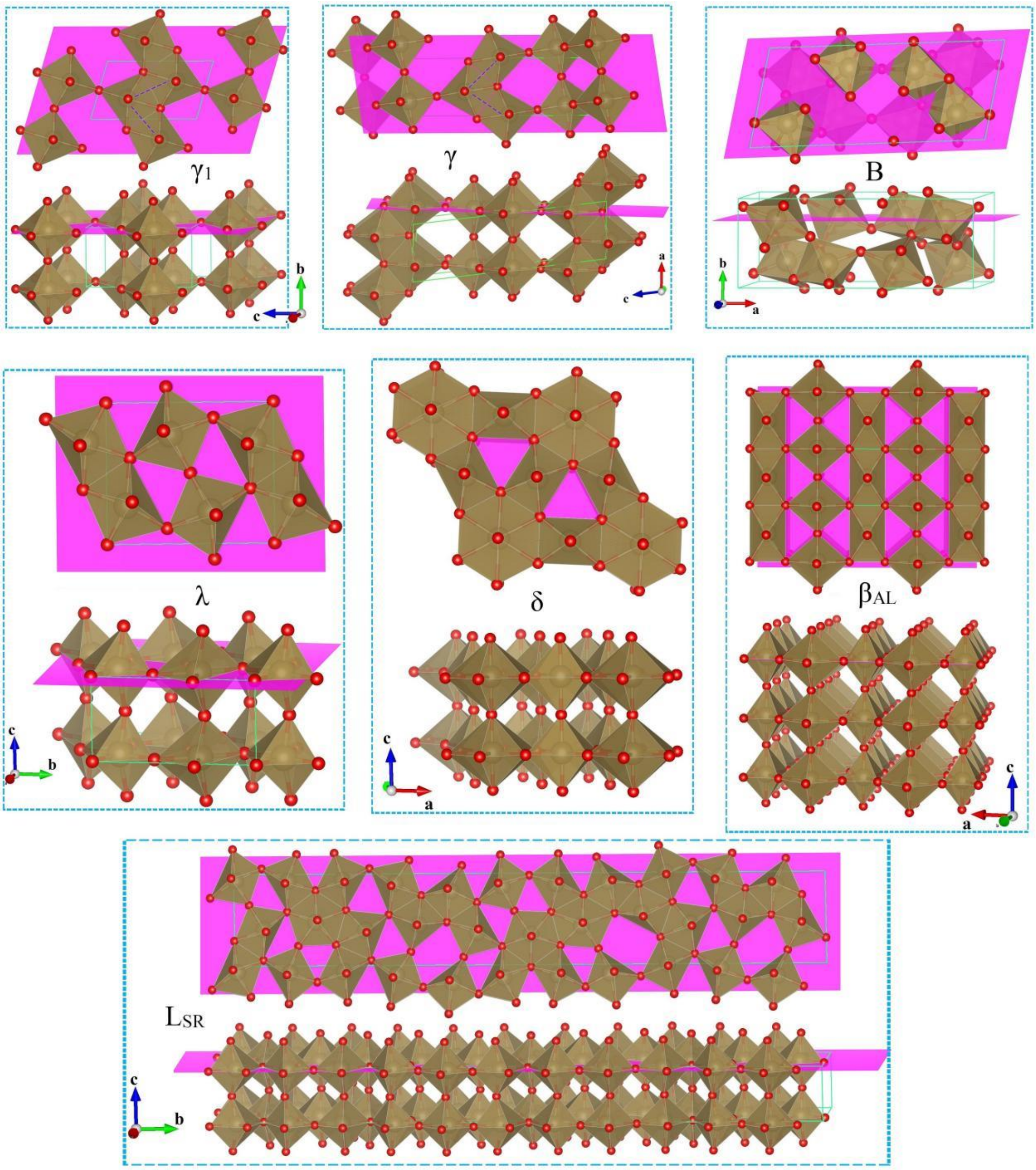

Fig. 2. Space arrangement of the $TaO_n$ polygons in a number of $Ta_2O_5$ crystal phases. For each phase, the basal plane showing the quasi-2D arrays of Ta is in upper panel.

**Table 3**. Geometries of the elementary building blocks $TaO_n$ in a number of $Ta_2O_5$ phases. The coordination number CN($TaO_n$) defined by Ta-Ta distances, and the volumes of polygon $TaO_n$ are presented for comparison. For $TaO_6$ of $L_{SR}$-$Ta_2O_5$, the averaged volume and its standard deviation are presented.

| $Ta_2O_5$ phases | $TaO_n$ | CN($TaO_n$) | V[$TaO_n$] (Å$^3$) |
|---|---|---|---|
| $\gamma_1$ | $TaO_6$ (×2) | 7 | 10.385 |
| $\gamma$ | $TaO_6$ (×4) | 7 | 10.347 |
| B | $TaO_6$ (×8) | 8 | 10.682 |
| $\lambda$ | $TaO_6$ (×4) | 8 | 10.157 |
| $L_{SR}$ | $TaO_6$ (×19)<br>$TaO_7$ (×3) | $TaO_6$: 7, 8<br>$TaO_7$: 8 | $TaO_6$: 10.202 ± 0.129<br>$TaO_7$: 13.854, 13.786, 14.109 |
| $\delta$ | $TaO_6$ (×3)<br>$TaO_8$ (×1) | $TaO_6$: 8<br>$TaO_8$: 8 | 9.883 (×3)<br>19.118 (×1) |
| $\beta_{AL}$ | $TaO_6$ (×4) | 8 | 9.994 (×2)<br>11.173 (×2) |

The elastic constants, which provide information about the strength and structural stability of the material, reflect the stress and strain relationship in anisotropic medium. Understanding of the mechanical properties is crucial to its potential applications in biological implant coating materials [55]. Herein, the elastic coefficients and anisotropy of $\gamma_1$-$Ta_2O_5$ are studied based on DFT calculations. The mechanical stability of the crystal structure is demonstrated by elastic coefficients and phonon spectra.

We start with fitting the energy-volume (E-V) data to Tait, Vinet, Murnaghan, Birch-Murnaghan 3rd-order and 4th-order equations of states (EOSs) [56-60] by using the program VASPKIT [61]. As shown in Fig. 1(d), the fitted curve of Birch-Murnaghan 4th-order EOS is in the best agreement with the DFT data points, while the other curves almost coincide with each other. Therefore, we choose the Birch-Murnaghan 4th-order EOS (see supplemental material) [60] to describe the relationship between total energy and crystal volume. According to the fitting data, the optimal values of bulk modulus ($B_0$), the first pressure derivative ($B_0'$) and the second pressure derivative ($B_0''$) of the bulk modulus turn out to be 207.458 GPa, 8.156 and -0.122 GPa$^{-1}$, respectively. With the determined EOS, we can get the volume and energy values at different pressure points within a certain range.

A triclinic crystal has totally 21 independent elastic coefficients due to its low structural symmetry. The flexibility coefficients of are obtained by solving the inverse of the elastic coefficient matrix. The polycrystalline bulk modulus (*B*) and shear modulus (*G*) are calculated based on the Voigt-Reuss-Hill approximation [62-64] (see supplemental material):

$$B = (B_V + B_R)/2,\ G = (G_V + G_R)/2, \tag{3}$$

where the subscripts *V* and *R* represent the elastic modulus obtained by the method of Voigt and Reuss, which are the upper and lower bound of the modulus [65], respectively. From $B$ and $G$ obtained by the Hill model (Eq. (3)), the Young's modulus (*E*) and Poisson's ratio (*v*) can be calculated as follows:

$$E = \frac{9BG}{3B+G},\ \nu = \frac{3B-2G}{6B+2G}. \tag{4}$$

Using Eq. (3), the bulk modulus $B$ of $\gamma_1$-$Ta_2O_5$ is calculated to be ~ 201.54 GPa, in good agreement with the result of fitting the Birch-Murnaghan 4th-order EOS. The values of *G*, *E*, *v* are 49.102 GPa, 136.24 GPa and 0.387, respectively. The data are comparable with the results reported for different $Ta_2O_5$ phases [17].

The elastic property of most low-symmetry single crystals is anisotropy. To quantitatively evaluate the elastic anisotropy of single crystalline $\gamma_1$-$Ta_2O_5$, a universal elastic anisotropy index $A^U$ is adopted [66]:

$$A^U = 5\frac{G_V}{G_R} + \frac{B_V}{B_R} - 6. \tag{5}$$

When $A^U$ = 0, the single crystal is isotropic. The larger $A^U$ is, the higher the degree of anisotropy would be expected. The calculated $A^U$ value of $\gamma_1$-$Ta_2O_5$ is 4.078, which is larger than all the values of nearly one hundred crystals presented by Ref. [66] in the elastic anisotropy diagram. It indicates the strong elastic anisotropy of $\gamma_1$-$Ta_2O_5$. This is in qualitative agreement with previous study on a number of $Ta_2O_5$ model structures [17]. To study the mechanical stability of the crystal structure, we have calculated the six eigenvalues of the elastic coefficient matrix. All the eigenvalues are positive, indicating that triclinic $\gamma_1$-$Ta_2O_5$ is mechanically stable [67]. In the following, the dynamical stability of $Ta_2O_5$ has been further confirmed through phonon spectrum calculations based on DFPT [42].

### 3.2. Vibrational and Thermal Properties

Figure 3 shows the phonon dispersion and vibrational density of states (VDOS) of $\gamma_1$-$Ta_2O_5$ along some high symmetry q-points. There is no any imaginary frequency point in the phonon spectrum, indicating the dynamical stability of $\gamma_1$ phase. From the partial VDOS, it is seen that Ta atoms contribute mainly to the vibrational modes of $\omega < 300$ cm$^{-1}$, while O atoms contribute mainly to the vibrations with frequencies above 100 cm$^{-1}$. This is can be understood from the large difference of atomic masses between O and Ta: For the same bonding strength, the vibrational frequency of a harmonic oscillator is proportional to the inverse of square root of atomic mass. The low-frequency region of $\omega < 150$ cm$^{-1}$ corresponds to the vibrations of the $TaO_6$ polyhedron. The mid-frequency region of 150 cm$^{-1}$ $< \omega <$ 400 cm$^{-1}$ corresponds to the deformation motion of O atom relative to the central Ta atoms. The high-frequency region ($\omega > 400$ cm$^{-1}$) corresponds to the stretching motions of the Ta-O bonds [30].

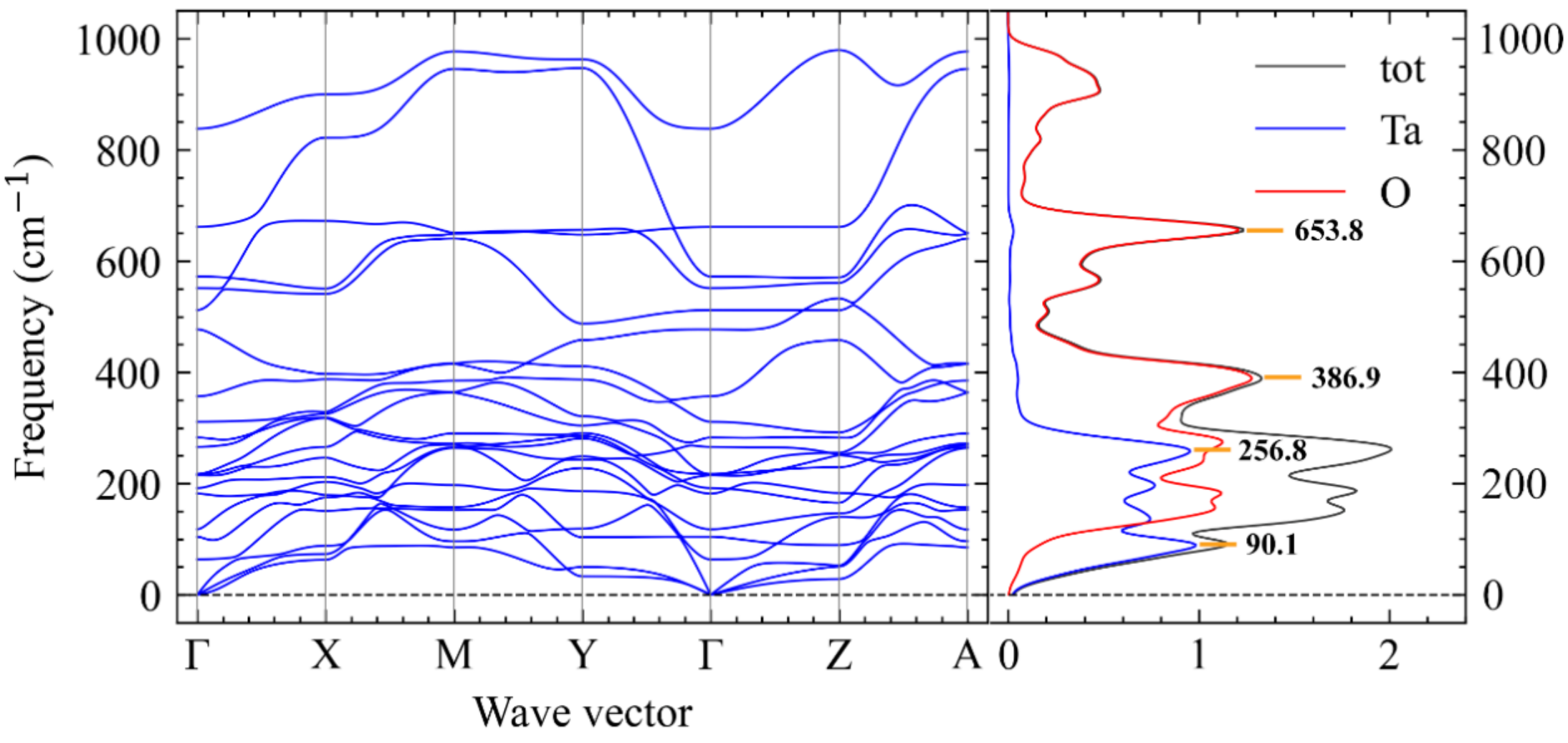

**Fig. 3.** Calculated phonon dispersion (left panel) along some high-symmetry lines of the q-points (wave vector) in the BZ and the VDOS (right panel) of $\gamma_1$-$Ta_2O_5$. The direct coordinates of the q-points in the BZ: Γ = (0, 0, 0), X = (0.5, 0, 0), M = (0.5, 0.5, 0), Y = (0, 0.5, 0), Z = (0, 0, 0.5), A = (0.5, 0.5, 0.5).

Based on the data of DFPT calculations, we are able to investigate the thermal properties of $\gamma_1$-$Ta_2O_5$ within the quasi-harmonic approximation (QHA) [43]. Heat capacity characterizes the ability of a material to absorb and release heat, which is mainly determined by lattice vibrations and the thermal motions of conducting electrons therein, and reflects the change of kinetic energy of microscopic particles with temperature. As a wide band gap semiconductor, the valence electrons of $Ta_2O_5$ are frozen near the Fermi level; there are very few conducting electrons at room temperature and below. Therefore, the main contribution to the heat capacity is the excitation due to phonons. For a single vibrational mode with an angular frequency $\omega_j$, the statistical average energy $\overline{E_j}(T) = \frac{1}{2}\hbar\omega_j + \frac{\hbar\omega_j}{e^{\hbar\omega_j/k_BT}-1}$; and the corresponding heat capacity $C_V^j$ at temperature $T$ is $C_V^j = \left(\frac{d\overline{E_j}(T)}{dT}\right)_V = k_B\left(\frac{\hbar\omega_j}{k_BT}\right)^2 \frac{e^{\hbar\omega_j/k_BT}}{\left(e^{\hbar\omega_j/k_BT}-1\right)^2}$. The constant-volume heat capacity ($C_V$) due to lattice vibrations/phonons is therefore the sum of all the vibrational modes: $C_V = \left(\frac{\partial E}{\partial T}\right)_V = \sum_j^{3N} C_V^j$, where $N$ is the number of atoms in the crystal, and $k_B$ and $\hbar$ are the Boltzmann constant and reduced Planck constant, respectively.

Figure 4 shows the variation of $C_V$ with temperature under QHA [43]. When the temperature is below 20 K, $C_V$ is proportional to $T^3$, which is consistent with the Debye model. From 20 to 750 K, the heat capacity increases rapidly and the rate of increase of $C_V$ gradually decreases with increasing temperature. At temperatures above 1000 K, $C_V$ converges to the classical thermodynamic limit (~ 172.969 J mol$^{-1}$ K$^{-1}$), which is consistent with the Dulong-Petit law ($C_V$ = $3Nk_B$ ~ 174.518 J mol$^{-1}$ K$^{-1}$).

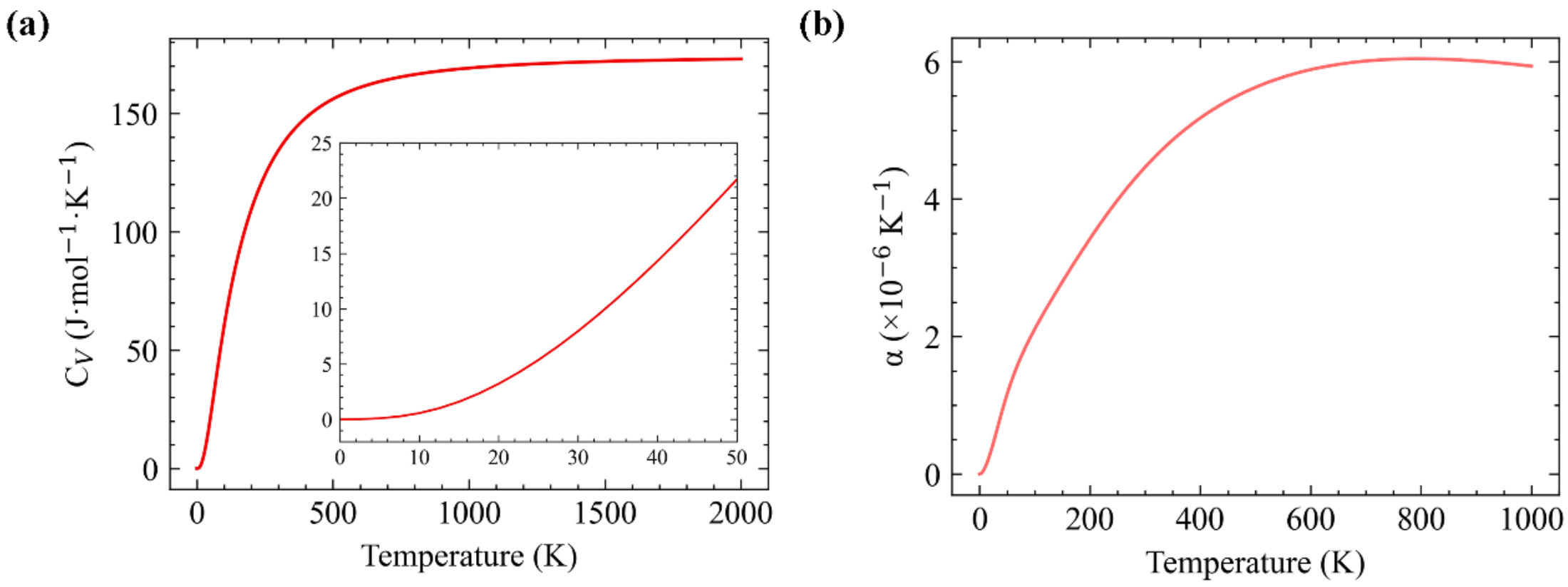

**Fig. 4.** (a) Heat capacity at constant volume ($C_V$) and (b) Thermal expansion coefficient (α) of $\gamma_1$-$Ta_2O_5$ as a function of temperature.

Thermal expansion is due to the cubic and higher order terms of interatomic interaction potentials, which manifest as the intensification of the nonlinear vibration of atoms around their equilibrium positions when temperature increases. The thermal expansion behavior of $\gamma_1$-$Ta_2O_5$ at finite temperature is evaluated by using QHA which takes into account the anharmonic effects. With the increase of temperature, the thermal expansion coefficient and heat capacity show a similar trend. It is worth noting that after reaching a maximum at ~ 750 K, the thermal expansion coefficient decreases slightly with temperature. According to previous experimental studies on $LaCoO_3$ [68], the appearance of such a maximum of thermal expansion coefficient is due to temperature-driven spin state transition. Therefore, the appearance of maximum thermal expansion coefficient at ~ 750 K may signify possible structural transition in $\gamma_1$-$Ta_2O_5$.

### 3.3. Electronic Properties

As seen from Table 1, the arrangement of atoms in real space has significant impacts on the band gap of $Ta_2O_5$. We went further to study the electronic properties of $\gamma_1$-$Ta_2O_5$ by carrying out first-principles calculations on its electron energy bands and density of states. Although the Kohn-Sham eigenvalues obtained by standard

DFT calculations are often taken as the electron excitation energies, they cannot accurately describe the energy change of electron removal or addition to the system. For the excited states of many-electron systems, the calculated band gaps of semiconductors or insulators by standard DFT are generally 30%-50% smaller than the experimental values [69, 70]. In the *GW* approximation, the one-particle Green's function can well describe the generation and annihilation of quasiparticles in the process of propagation, and the screening effects on the Coulomb interactions *W* are taken into account. Based on the results of DFT-GGA, we use the *GW* method as described above to perform quasiparticle corrections to obtain a more accurate band structures.

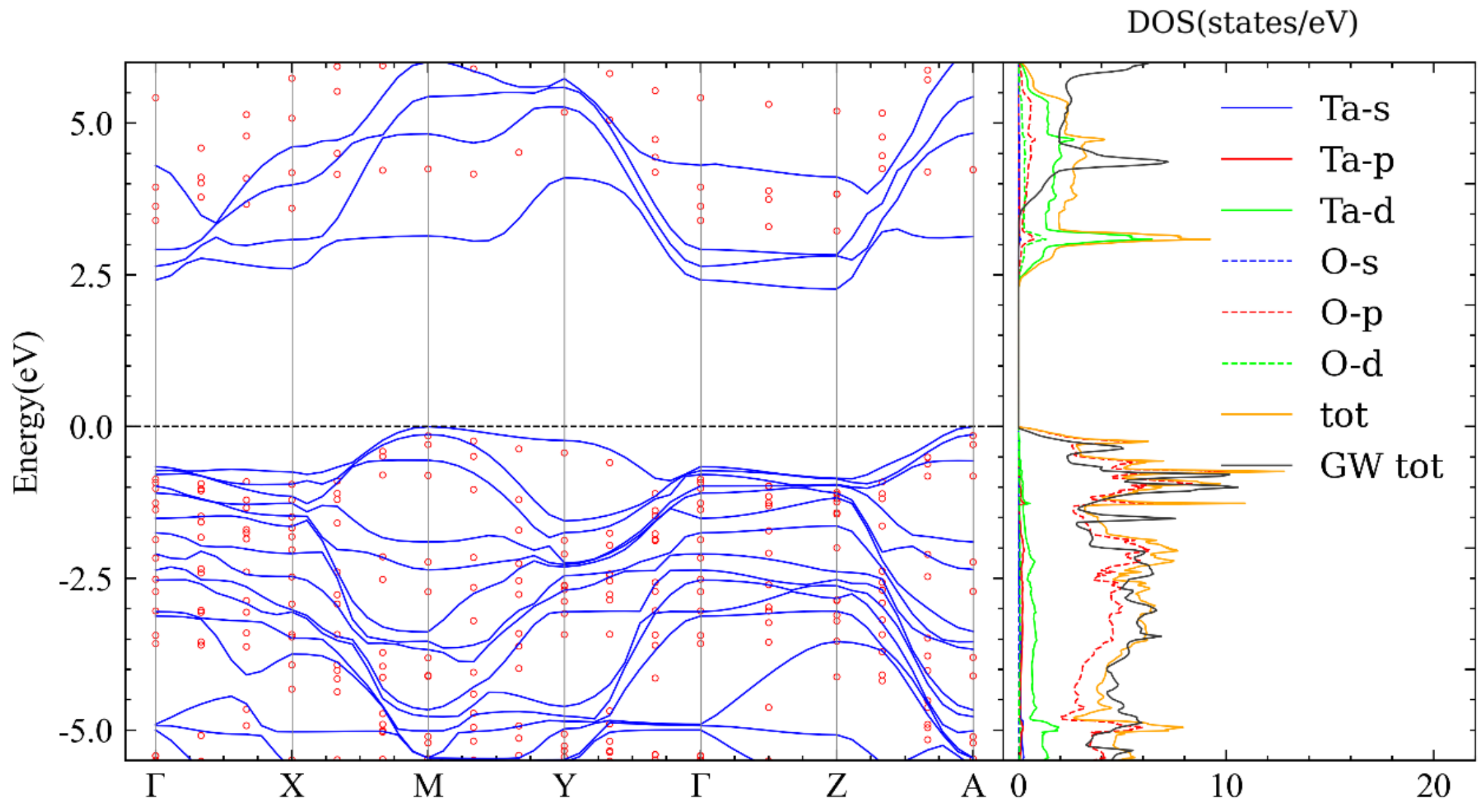

**Fig. 5.** Band structures, density of states (DOS), partial density of states (PDOS) of $\gamma_1$-$Ta_2O_5$ calculated by DFT-GGA and *GW* method. In the band diagram, the blue solid line and scattered red circles are the results of GGA and *GW* along the lines of high-symmetry k-points, respectively. PDOS are the results of GGA. The direct coordinates of k-points in the BZ: Γ = (0, 0, 0), X = (0.5, 0, 0), M = (0.5, 0.5, 0), Y = (0, 0.5, 0), Z = (0, 0, 0.5), A = (0.5, 0.5, 0.5).

We focus on the electronic structures near the Fermi level, which determine excited properties of the material. Figure 4 shows the calculated band structures and density of states of $\gamma_1$-$Ta_2O_5$ along the high symmetry k-point lines of BZ. The top of the valence band is located at points M and A, and the bottom of the conduction band is located at the Z point, which indicates that $Ta_2O_5$ is an indirect band gap semiconductor with a GGA band gap of 2.238 eV. The direct band gap by GGA is at the Γ point with a value of 3.065 eV. After correction by the *GW* method, the indirect band gap is increased to ~ 3.361 eV and the direct band gap is still at the Γ point with a value of ~ 4.264 eV. Compared to the results of DFT-GGA, the valence band energy only decreases to a small extent while the conduction band rises significantly. Such a variation trend is in line with previous studies [45, 71]. The overall dispersion the entire band structure remains nearly unchanged. The valence band maximum (VBM) and the conduction band minimum (CBM) appear at the same k-points, but the band gap is substantially enlarged. As can be seen from the partial density of state (PDOS), the electronic states at the top of the valence band are mainly composed of O 2p orbitals, and the bottom of the conduction band mainly consists of Ta 5d states. Meanwhile, the steep PDOS of Ta 5d orbitals indicates the high localization of d-electrons near the bottom of conduction bands.

## 4. Conclusions

To summarize, we have identified a triclinic phase of $Ta_2O_5$ ($\gamma_1$-$Ta_2O_5$) with one formula unit (Z=1) at atmospheric pressure by using structure search based on PSO algorithm and first-principles calculations. The structural, elastic, phonon, thermodynamic and electronic properties of $\gamma_1$-$Ta_2O_5$ are systematically investigated. According to ground-state energies, $\gamma_1$-$Ta_2O_5$ is the most stable among the Z=1 $Ta_2O_5$ structures, including the original structures reported by the Materials Project (MP) [53]. Its dynamical stability has been confirmed by phonon spectrum. Through structural refinement, we show that $\gamma_1$-$Ta_2O_5$ is actually isomorphic to two MP structures (MP-1, MP-2), and the third MP structure (MP-3) is actually the building block of the experimentally identified $\beta_{AL}$ phase [52]. Furthermore, our DFT-*GW* calculations predict $\gamma_1$-$Ta_2O_5$ is a wide band gap semiconductor with an indirect gap of ~ 3.361 eV. The Z=1 structural models of $Ta_2O_5$ studied in this work can serve as the elementary building blocks for constructing new low-temperature phases of $Ta_2O_5$ at atmospheric pressure. Its combination with the other previously reported phases has guiding significance for the synthesis of far more complex $Ta_2O_5$ crystal phases and the experimental preparation of polycrystalline $Ta_2O_5$ with adjustable band gaps and optical properties.

## 5. Supplemental Material

Details of the 4th-order Birch-Murnaghan EOS [60] and the Voigt-Reuss-Hill approximation [62-64] are available online.

## Acknowledgements

This work is financially supported by the National Natural Science Foundation of China (No. 12074382, 11804062, 11474285). We are grateful to the staff of Hefei Branch of Supercomputing Center of Chinese Academy of Sciences, and Hefei Advanced Computing Center for their support of supercomputing facilities.